\documentclass[showpacs,onecolumn]{revtex4-1}

\usepackage{graphicx}
\usepackage{dcolumn}
\usepackage{amssymb,amsmath}
\usepackage{natbib}
\usepackage{subfigure}
\usepackage{epstopdf}
\usepackage{placeins}
\usepackage{color}
\usepackage{setspace}

\begin{document}

\title{Flow states in two-dimensional Rayleigh-B\'enard convection as a function of aspect-ratio and Rayleigh number}
\author{Erwin P. van der Poel$^{1}$, Richard J. A. M. Stevens$^{1}$, Kazuyasu Sugiyama$^{2}$, and Detlef Lohse$^{1}$}
\affiliation{
$^{1}$Department of Physics, Mesa+ Institute,  and J.\ M.\ Burgers Centre for Fluid Dynamics, University of Twente, 7500 AE Enschede, The Netherlands \\
$^{2}$Department of Mechanical Engineering, School of Engineering, The University of Tokyo, Tokyo, Japan}
\date{\today}

\begin{abstract}
In this numerical study on two-dimensional Rayleigh-B\'enard convection we consider $10^7 \leq Ra \leq 10^{12}$ in aspect ratio $0.23 \leq \Gamma \leq 13$ samples. We focus on several cases. First we consider small aspect ratio cells, where at high $Ra$ number we find a sharp transition from a low $Ra$ number branch towards a high $Ra$ number branch, due to changes in the flow structure. Subsequently, we show that the influence of the aspect ratio on the heat transport decreases with increasing aspect ratio, although even at very large aspect ratio of $\Gamma\approx10$ variations up to $2.5\%$ in the heat transport as a function of $\Gamma$ are observed. Finally, we observe long-lived transients up to at least $Ra=10^9$, as in certain aspect ratio cells we observe different flow states that are stable for thousands of turnover times.
\end{abstract}

\maketitle
\section{Introduction}

In Rayleigh-B\'enard (RB) convection \cite{ahl09,loh10} a fluid in a box is heated from below and cooled from above. This system is paradigmatic for turbulent heat transfer, with many applications in atmospheric and environmental physics, astrophysics, and process technology. Its dynamics are characterized by the Rayleigh number $Ra = \beta g \Delta L^3/(\kappa\nu)$, the Prandtl number $Pr = \nu /\kappa$ and the aspect-ratio $\Gamma = d/L$. Here, $L$ is the height of the sample and $d$ its width, $\beta$ is the thermal expansion coefficient, $g$ the gravitational acceleration, $\Delta$ the temperature difference between the bottom and the top of the sample, and $\nu$ and $\kappa$ the kinematic viscosity and the thermal diffusivity, respectively. 

To better understand geo- and astrophysical phenomena it is important to know the heat transfer in the high $Ra$ number regime. Nowadays, most experimental and numerical measurements of the heat transfer, indicated by the Nusselt number $Nu$, in 3D agree up to $Ra\approx 2\times 10^{11}$ (see Ref. \cite{ahl09} for detailed references) and these data are well described by the Grossmann-Lohse (GL) theory \cite{gro00,gro01,gro02,gro04}. Unfortunately, there are considerable unexplained differences in the heat transfer measurements for higher $Ra$ \cite{cha01,nie00,nie01,nie06,nie10b,roc01,roc02,roc10,fun09,ahl09c,ahl09d,ahl11,he11,ste10,ste10d}.

Explaining the differences between the high $Ra$ number measurements is subject of an ongoing discussion \cite{ahl09,roc10,nie10b,ste10,ste10d}. The difficulty is that experiments in the high $Ra$ number regime are very challenging as  it requires very large setups that operate under high pressures \cite{fun09,ahl09c,ahl09d,ahl11,he11}, or use liquid helium near its critical point \cite{cha01,nie00,nie01,nie06,roc01,roc02,roc10}. Therefore only a few setups in the world can perform high $Ra$ number experiments and testing the influence of some effects, e.g.\ the influence of the aspect ratio, is very time consuming \cite{roc10}. In 3D simulations it is difficult to study the problem, as these simulations are very CPU-time intensive \cite{ste10d}.

\begin{figure*}
\subfigure[]{\includegraphics[width=0.54\textwidth]{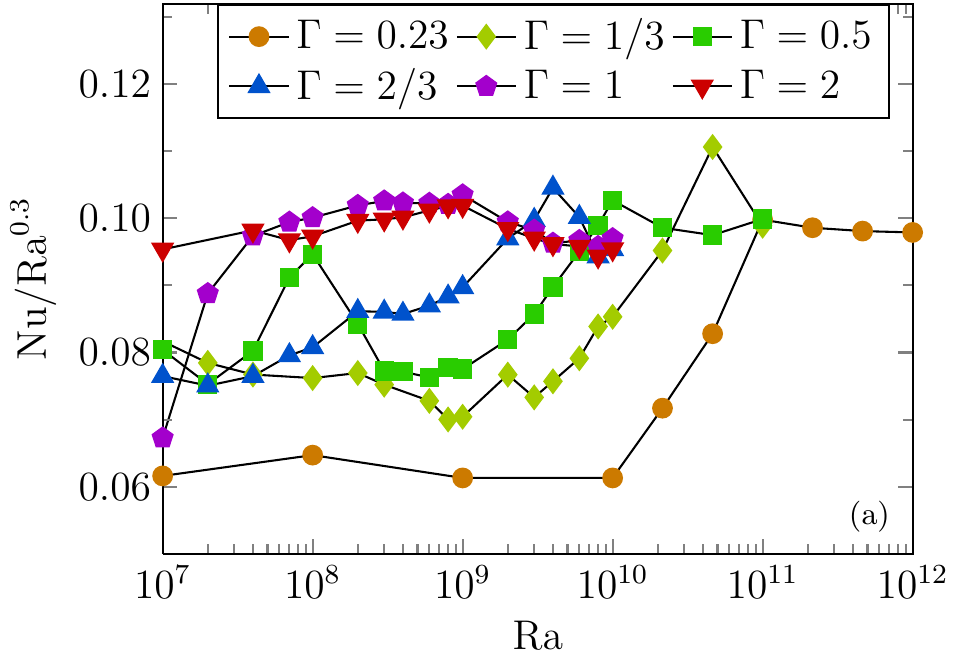}}
\subfigure[]{\includegraphics[width=0.40\textwidth]{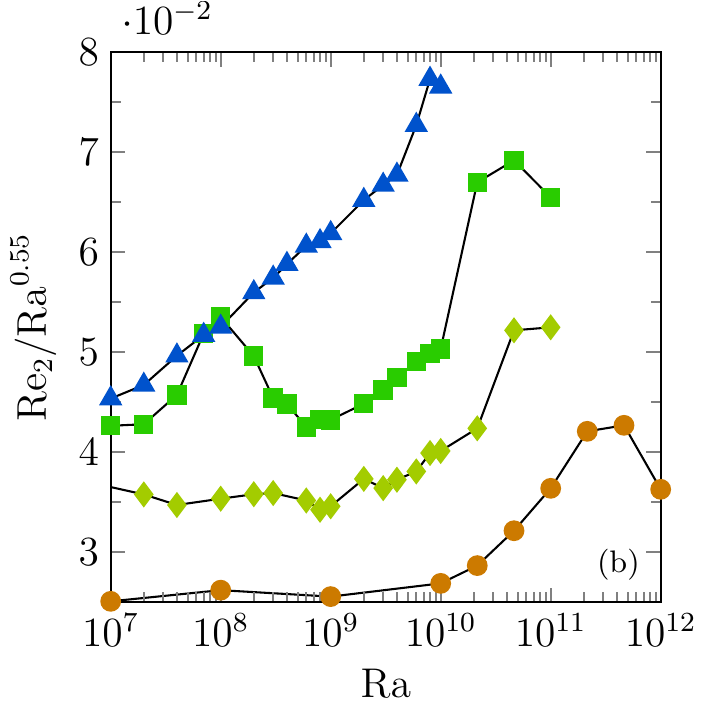}}
    \caption{a) $Nu$ versus $Ra$ for $Pr=1$ in aspect ratio $\Gamma=0.23$ (orange circles), $\Gamma=1/3$ (lime diamonds), $\Gamma=1/2$ (green squares), $\Gamma=2/3$ (blue upward pointing triangles), $\Gamma=1$ (purple pentagrams) and $\Gamma=2$ (red downward pointing triangles) samples. b) $Re_{2}$ versus $Ra$ for $Pr = 1$ and in $\Gamma < 1$ samples. The averaging time ranges from $4000~t_E$ to $200~t_E$.  }
\label{figure1}
\end{figure*}

\begin{figure}
\includegraphics[width=0.45\textwidth]{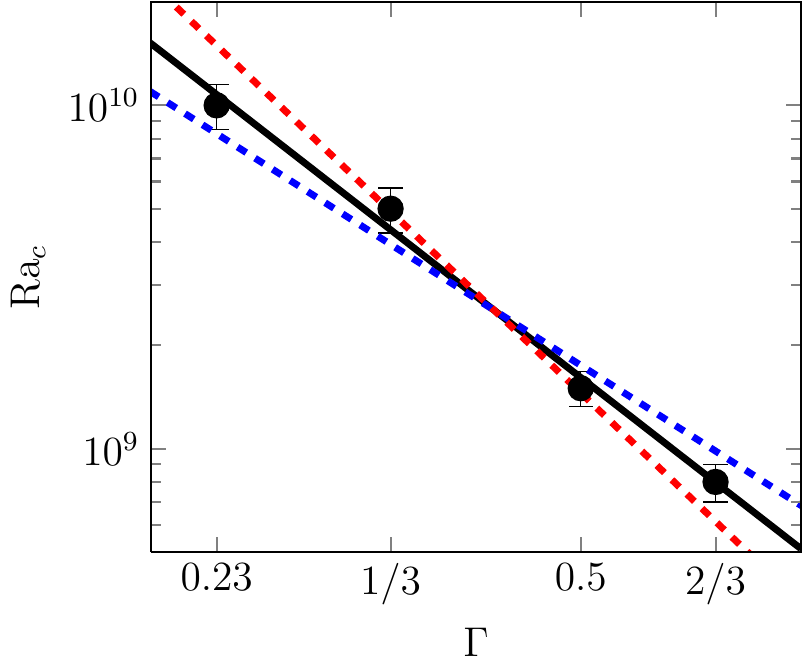}
    \caption{$Ra_c$ versus $\Gamma$, both on logarithmic scale. A least-squares fit, which is indicated by the black solid line, gives $Ra_c \sim \Gamma^{-2.44}$. The blue and red dashed lines give $Ra_c \sim \Gamma^{-2}$ and $Ra_c \sim \Gamma^{-3}$, respectively. }
\label{figure2}
\end{figure}

One possibility to limit the required computational time is to study the influence of several effects in 2D RB convection in order to select the most interesting cases that need to be studied in 3D. One may wonder whether the dynamics of 3D RB convection are sufficiently captured in 2D simulations. This question has been addressed by Schmalzl {\it et al.} \cite{sch04} and they showed that in relatively large aspect ratio cells, where several convection rolls can form next to each other, and for $Pr \gtrsim 1$ 2D RB convection global quantities such as the $Nu$ number and the $Re$ number show similar behavior in 3D. Therefore 2D simulations have been used to investigate a number of aspects observed in 3D convection \cite{sug07,ahl08,sug09,zho10,zho10b,zho11,ste12,joh09,ste10d}. 

It has been argued that the differences between high $Ra$ number experiments are due to the formation of different turbulent states. Experimental indications for coexistence of different turbulent states in RB convection are found by several authors \cite{ roc02,chi04a,sun05a,xi08,wei10b,nie10b,ahl09d,ahl11} and are discussed theoretically by Grossmann and Lohse \cite{gro11}. It has been shown by Xi $\&$ Xia \cite{xi08} and Weiss $\&$ Ahlers \cite{wei10b} that in a $\Gamma=0.5$ sample the flow can be either in a single roll state (SRS) or in a double roll state (DRS), each with a specific heat transport. Later van der Poel {\it et al.} \cite{poe11} confirmed this effect in 2D simulations and showed that the flow state, and thus $Nu$, can strongly depend on the aspect ratio. Experiments by Roche {\it et al.} \cite{roc10} show that the transition towards a regime in which a strong heat transfer enhancement as function of $Ra$ is found strongly depends on the aspect ratio and the experiments by Niemela $\&$ Sreenivasan \cite{nie10b} reveal the presence of a low and a high $Ra$ number branch. 

Using a direct numerical simulation scheme for $10^7 \le Ra \le 10^{12}$ we show that also in small aspect ratio 2D RB convection there is a low and a high $Ra$ number branch for the heat transport and that the transition at the critical Rayleigh number $Ra_c$ between the branches increases with decreasing aspect-ratio. Subsequently, we study the heat transport as function of the aspect ratio at fixed $Ra$. Bailon-Cuba {\it et al.} \cite{bai10} have studied the effect on $Nu$ of increasing the aspect-ratio in 3D numerics and found that $Nu$ depends on the number of horizontally stacked rolls. Now, we try to estimate how large the aspect ratio should be before finite size effects become unimportant.

The objective of this paper is to study the phenomena associated with varying $Ra$ and $\Gamma$ to a large extent. The corresponding observations of the double branching in $Nu(Ra)$ and vanishing variations in $Nu(\Gamma)$ at large $Ra$ and $\Gamma$ can be found in chapters III and IV, respectively.
Furthermore,  in chapter V we will discuss some long-lived transient effects we observe at $Ra = 10^9$ and $Pr = 0.7$. These three cases have overlapping origin and secondary effects, such as multi-stability and roll break-up.

In the case of long-lived transients, the exhibited multiple roll states with different corresponding $Nu$ and $Re$, survive for such long times, that the roll state can either be interpreted as a transient or as stable. This classification is very difficult, as the insuperable finite run time of simulations makes it impossible to ascertain that the observed state is not a transient. To overcome the ambiguity between transient and multi-stability, the cases for which this classification is unclear will be termed as transients and these will be annotated with the simulation run time. This time is an indication of the stability of the observed states.
We conclude the paper with a discussion of the presented results.

\begin{figure*}
\subfigure[]{\includegraphics[width=0.19\textwidth]{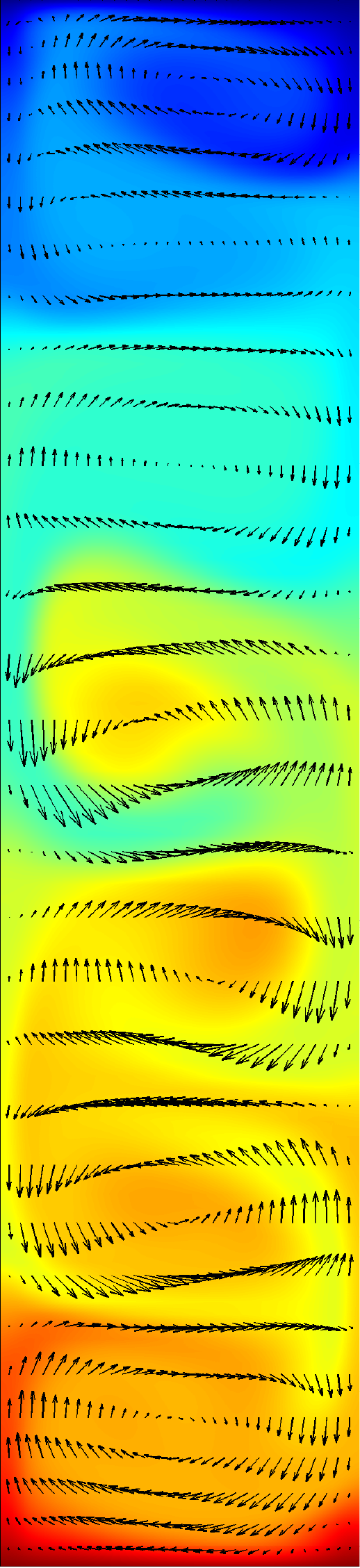}}
\subfigure[]{\includegraphics[width=0.19\textwidth]{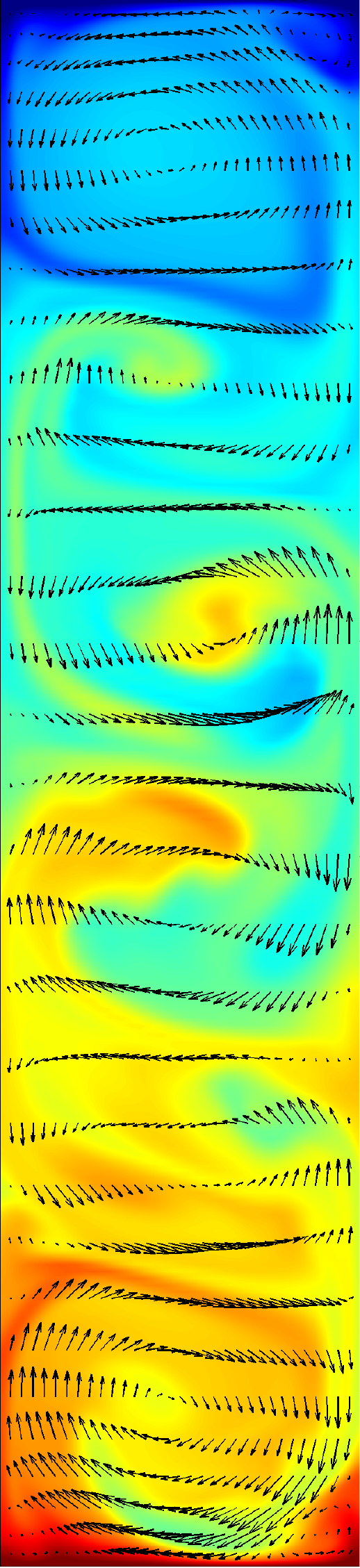}}
\subfigure[]{\includegraphics[width=0.19\textwidth]{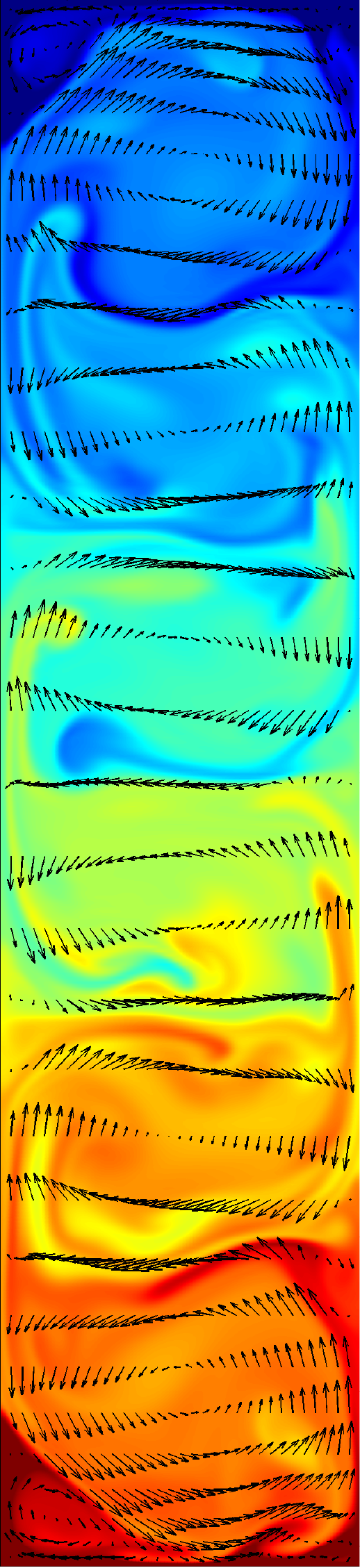}}
\subfigure[]{\includegraphics[width=0.19\textwidth]{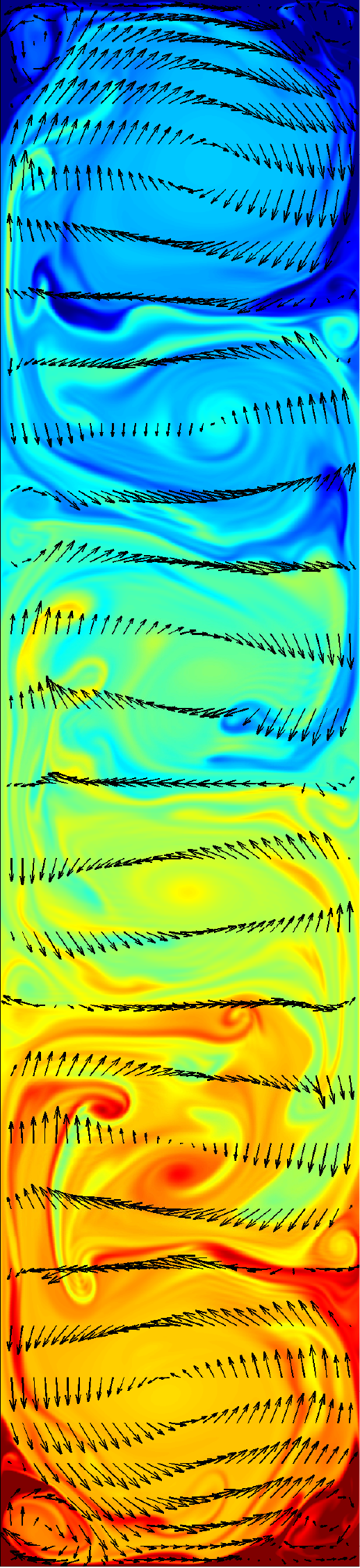}}
\subfigure[]{\includegraphics[width=0.19\textwidth]{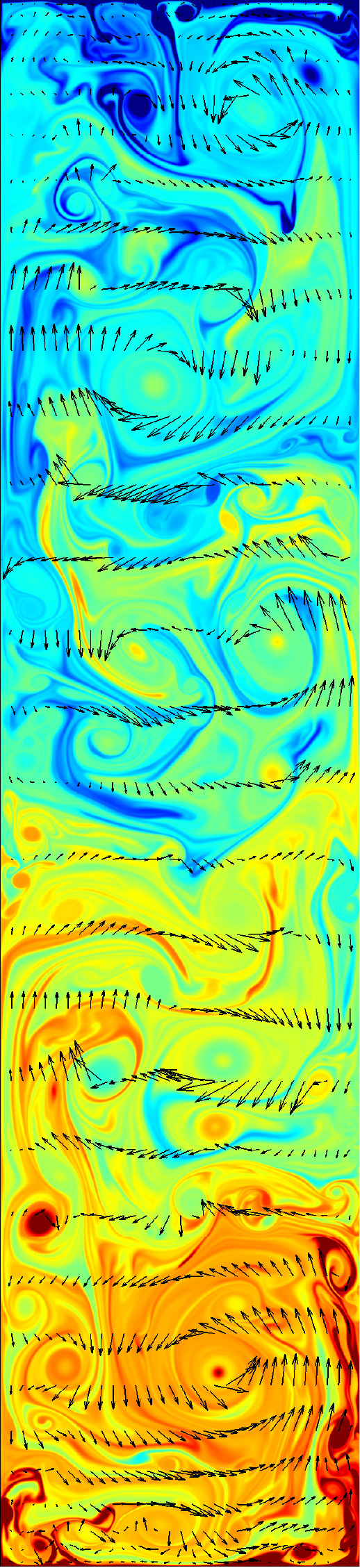}}

    \caption{Flow visualisation for the flow in a $\Gamma=0.23$ sample with $Pr = 1$ at (a) $Ra=10^8$, (b) $Ra=10^9$, (c) $Ra=10^{10}$, (d) $Ra=10^{11}$, and (e) $Ra=10^{12}$. The velocity field is represented by the vectors and, red and blue indicate high and low temperature, respectively. Note that the colormap is not constant. For $Ra=10^{11}$ the activity of the rolls is increased and at $Ra=10^{12}$ the roll structure appears to be completely broken down. Flow field movies corresponding to the sharp transition can be found in the supplemental material.}
\label{figure3}
\end{figure*}

\section{Numerical method} \label{Section_NumericalMethod}

The code on which the results in this paper are based is a fourth order finite difference discretization of the incompressible Oberbeck-Boussinesq equations. The velocity boundary conditions on the walls are no-slip. The (relative) temperature is fixed at $\Delta/2$ at the bottom plate and $-\Delta/2$ at the top plate, with adiabatic sidewall conditions. The code has been described and tested in detail by Sugiyama {\it et al.} \cite{sug07,ahl08,sug09} and this code was used to explain the non-Oberbeck Boussinesq effects observed in 3D experiments of Ahlers {\it et al.}\ \cite{ahl08}, and Zhou {\it et al.} \cite{zho10b,zho11} and Sugiyama {\it et al.} \cite{sug10} found a good agreement between results obtained using this 2D code and experiments performed in quasi-2D RB samples.  The grid resolution we used obeys the strict criteria formulated in refs.\ \cite{ste10,shi10}. The Nusselt number $Nu$ is calculated with four different procedures. The $Nu$ used in this study is the average of
\begin{equation}
Nu = -\frac{\langle \partial_z \theta \rangle_{A,t}}{\Delta L^{-1}}
\end{equation}
evaluated at the top and bottom plate. Here $\langle \cdot \rangle_{A,t}$ denotes the average over a horizontal line and time. The resulting $Nu$ is checked using the relation between the volume-averaged dissipation rates and $Nu$ \cite{shr90}, given by
\begin{equation}
Nu = \frac{L^4}{\nu^2}Ra^{-1}Pr^2 \langle (\partial_iu_i)^2 \rangle_{V,t} + 1,
\end{equation}
\begin{equation}
Nu = \frac{L^2}{\Delta^2} \langle (\partial_i\theta)^2 \rangle_{V,t},
\end{equation}
where $u$ is the velocity normalized with the free fall velocity $U = \sqrt{\beta g \Delta L}$, $\theta$ is the dimensionless temperature $0 \le \theta \le 1$ and $\langle \cdot \rangle_{V,t}$ denotes the average over the complete domain and time. As the Nusselt number should not be dependent on the procedure by which it is calculated, comparing the results of the different procedures is an indication of the numerical error. For all data points in this study the error calculated by comparing the results is smaller than dot size. The time convergence of $Nu$ is checked by comparing the time-averages over half the duration of the simulation and the full duration. Again, the resulting error is for most cases smaller than dot size. However, this error is not necessarily monotonically decreasing with increasing simulation time due to slow trends and rare events. For an indication of the uncertainty associated with the calculation of the numerical quantities, the typical averaging time will be annotated with the data. The used unit of time is the large eddy turnover time, given by $t_E\equiv 4\pi/\langle|\omega_c|\rangle_t$, where $\omega_c$ is the vorticity at the center of a roll.

\section{High Rayleigh number convection} \label{Section_highRa}

Figure \ref{figure1}a shows the heat transport as function of $Ra$ for different aspect ratio cells. Just as in the experiments of Niemela $\&$ Sreenivasan \cite{nie10b} a low and a high $Ra$ number branch is observed and the $Nu$ difference is larger for the low aspect ratio cells. One can see that the $Ra$ number at which the transition from the low to the high branch sets in increases with decreasing aspect ratio. To find the transitional $Ra$ number $Ra_c$ we extrapolated the low branch, which scales approximately as $Nu \thicksim Ra^{0.30}$, and fitted the points found in the transition region with a power law. The intersection of the resulting lines gives $Ra_c$, which is calculated for $0.23 \leq \Gamma \leq 2/3$. Figure \ref{figure2} shows that $Ra_c$ scales approximately as $Ra_c \sim \Gamma^{-2.44}$, although the $\Gamma$-range is of course very limited. This is very similar to the result obtained by Roche {\it et al.} \cite{roc10}, who performed high $Ra$ number experiments in different aspect ratio samples and observed a sharp increase in the heat transfer, which sets in beyond a critical $Ra$ that scales as $Ra_c \sim \Gamma^{-2.50}$. However, we note that both data sets would also be in reasonable agreement with a $\Gamma^{-3}$ scaling, which would be equivalent to saying that a $Ra$ number based on the diameter $d$ and not the height $L$ of the cell is constant \cite{roc10}.

The transitional $Ra$ numbers we find are well below the value at which the transition towards the ultimate state is expected \cite{ahl09} and therefore it is important to know what effect is responsible for the sharp increase in the heat transport. To answer this question we visualized the flow field for the simulations in the aspect ratio $\Gamma=0.23$ sample, where the transition is most pronounced. Figure \ref{figure3} reveals that for low $Ra$ there are vertically stacked stable rolls. The flow structure is substantially different in the higher $Ra$ number cases. For $Ra=1\times10^{11}$ the increased temperature gradients in the corner rolls indicate that they are more active than at lower $Ra$, i.e. the corner rolls exchange more thermal energy with the large scale flow. As this increases the effective plate area that contributes to the heat transport, this leads to heat transport enhancement. For even higher $Ra$ the rolls (partially) break up due to increased buoyancy forces. The plumes break up the rolls and at the same time have a decreased travel distance towards the opposite plate. This leads to a higher heat transport. In order to verify this picture we calculated $Re_{2}$, where $Re_{i}\equiv u^{RMS}_{i}L/\nu$ is the Reynolds number based on the root mean square velocity with the index $i = 1,2,a$ indicating horizontal, vertical and absolute velocity, respectively. Figure \ref{figure1} shows that the transition in the $Nu$ number coincides with a strong increase in the vertical velocities. Therefore we can conclude that the transition we see here is due to a transition between different turbulent states - the breakdown of the corner-roll - and is not a transition towards the ultimate regime.

When we compare the 2D results with the measurements of Roche {\it et al.} \cite{roc10} we notice that the $Ra$ number at which the sharp increase in the heat transfer is observed happens at a lower $Ra$ in 2D than in 3D. Another difference is that we observe a low and a high $Ra$ number branch for the heat transfer in 2D, whereas no high $Ra$ number branch is observed in the experiments of Roche {\it et al.} \cite{roc10}. However, we note that Niemela $\&$ Sreenivasan \cite{nie10b} did observe two branches and it could well be that this second branch could also be found in the Roche {\it et al.} \cite{roc10} experiments at higher $Ra$. Although one should be careful with comparing the results from 2D and 3D RB convection in small aspect ratio samples, as we will discuss in detail below, we emphasize that it is important to keep in mind is that not every increase in the scaling of the heat transport as function of $Ra$ that is observed indicates a transition towards the ultimate regime, as we showed here for the 2D case.

Again we stress that one should be careful with a direct comparison between 2D and 3D results in small aspect ratio cells. For small aspect ratio cells the dynamics in 2D and 3D RB convection can be different as vertically stacked rolls are formed. In 2D vertically stacked rolls can lead to a larger reduction of the heat transport than in 3D RB convection as warm (cold) fluid in the bottom (top) part of the cell is more likely to be trapped than in 3D. In fact experiments and simulations of 3D RB convection, e.g.\ Bailon-Cuba {\it et al.} \cite{bai10}, show a weaker aspect ratio dependence than observed in the 2D simulations (section \ref{Section_LargeAR}) .

\section{Large aspect ratio} \label{Section_LargeAR}

\begin{figure*}
	\centering
\includegraphics[width=0.99\textwidth]{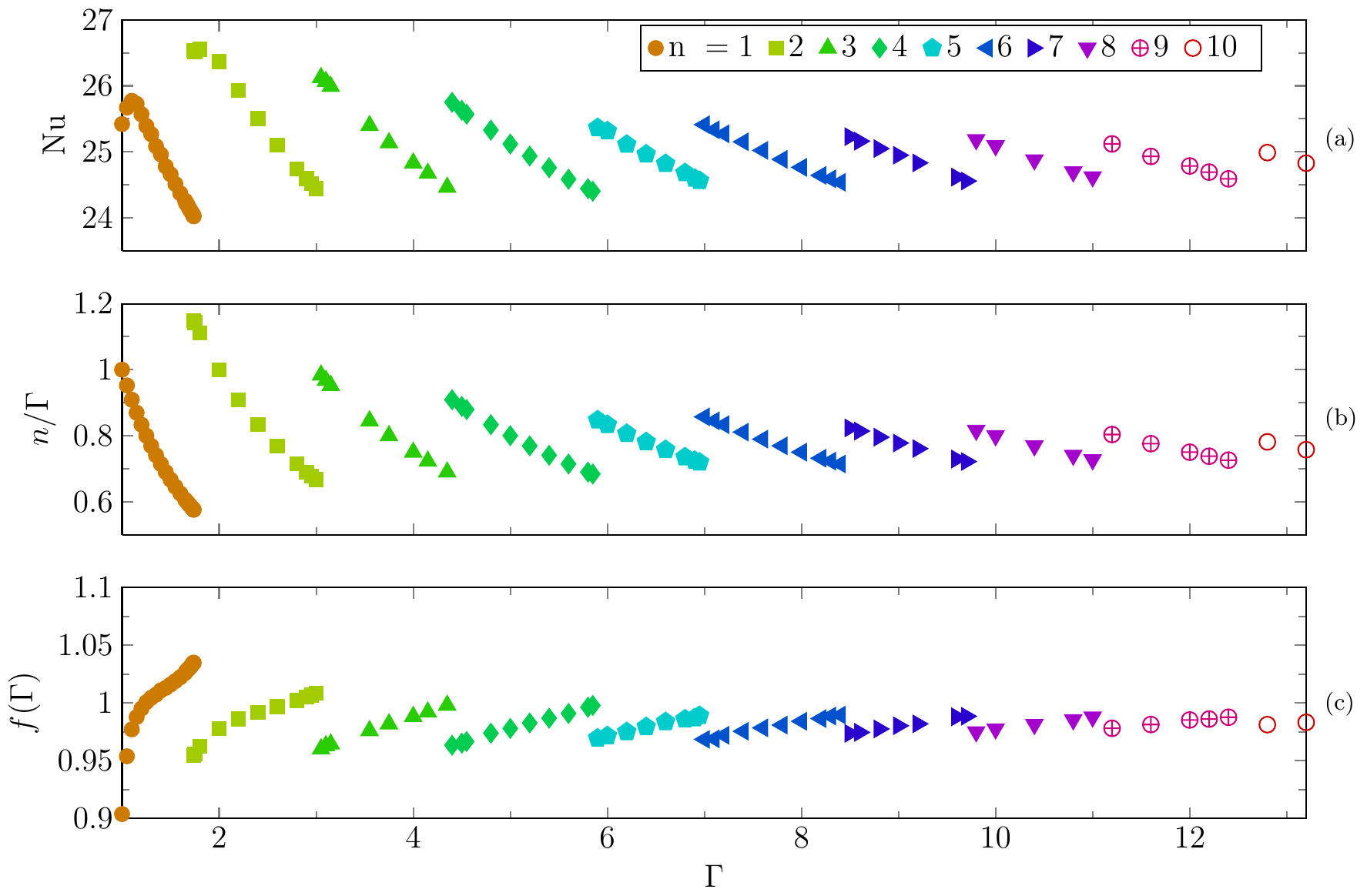}
    \caption{a) Nu as a function of $\Gamma$ for $Pr = 4.3$ and $Ra = 10^8$. The depicted $1 < \Gamma < 12$ range captures the SRS up to the horizontally stacked 10RS ($n = 10$), each represented by a cluster of points from low to high $\Gamma$.
    b) $n/\Gamma$ approaches a constant value of approximately $0.77$.
    c) The function $g(\Gamma)\equiv Re_2/(Re_1n/\Gamma)$. The error on the time-convergence is smaller than the dot size as the averaging time is more than $1000~t_E$.}
\label{figure4}
\end{figure*}

\begin{figure*}[ht]
	\centering
\includegraphics[width=0.88\textwidth]{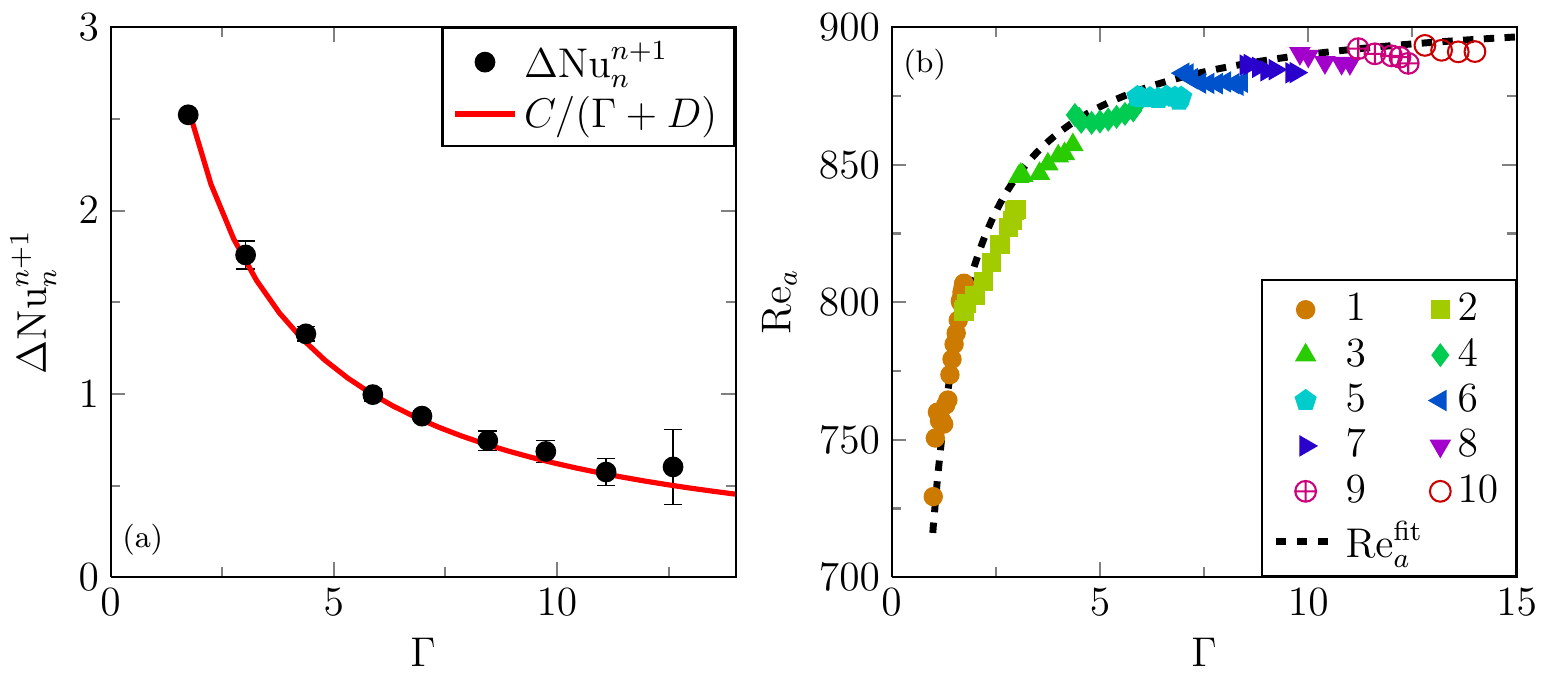}
    \caption{a) The amplitude of the jumps between different in roll states in $Nu(\Gamma)$, $\Delta\text{Nu}_n^{n+1}$ and the $C/(\Gamma+D)$ scaling as a function of $n$. Here the scaling constants $C=6.9$ and $D=1$. b) The Reynolds number $Re_{a}$ as a function of $\Gamma$ appears to asymptotically approach a constant $Re_{a,\infty}$ value at high $\Gamma$. The fit with equation \eqref{refit} is plotted as the dashed line.}
\label{figure5}
\end{figure*}

\begin{figure*}
	\centering
\includegraphics[width=0.88\textwidth]{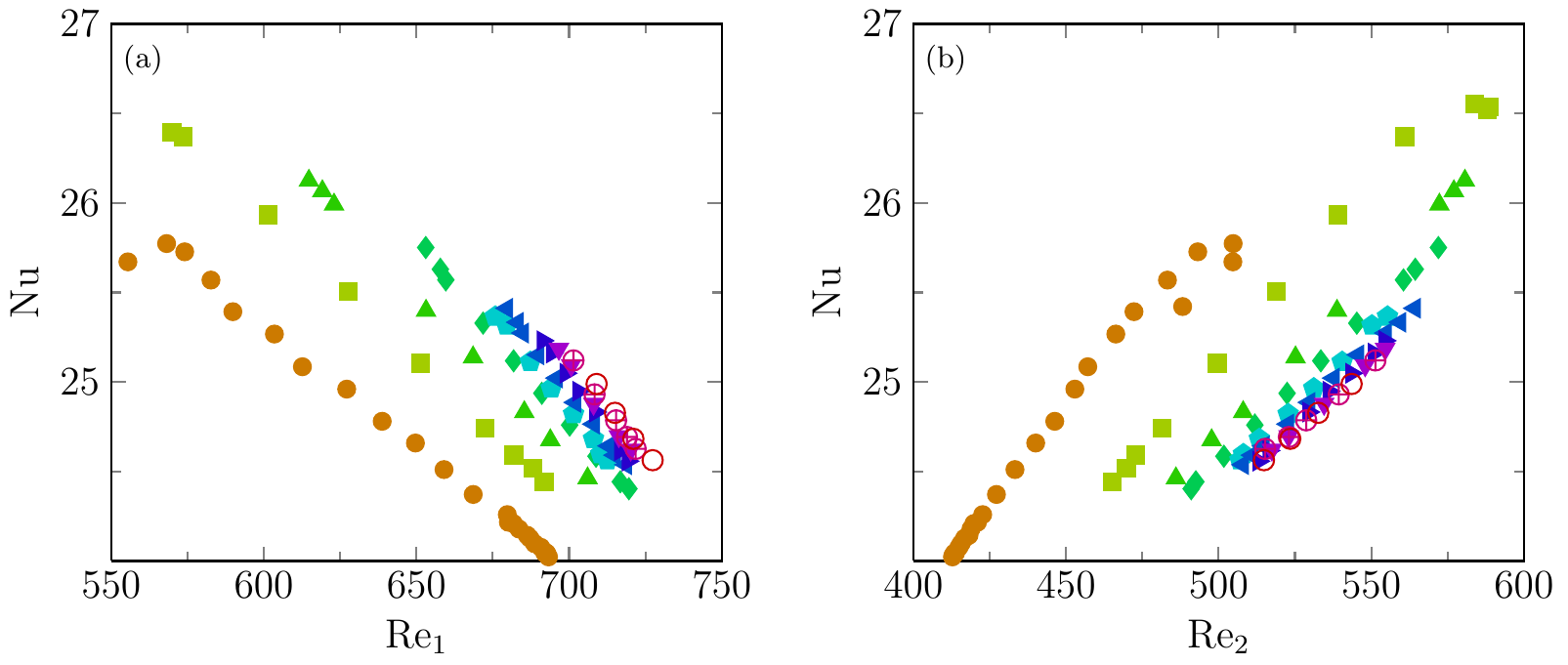}
    \caption{$Nu$ versus $Re_1$ (a) and $Re_2$ (b) for $Ra=10^8$ and $Pr=4.3$. Different $n$ are indicated by the colored symbols, see legend figure \ref{figure5}b. A negative correlation between $Nu$ and $Re_1$ and a positive correlation between $Nu$ and $Re_2$ can be seen.}
\label{figure6}
\end{figure*}

To find the critical horizontal length of the system at which the heat transport becomes nearly independent of the aspect ratio we extend the dataset for $Ra = 10^8$ and $Pr = 4.3$ of van der Poel {\it et al.} \cite{poe11} towards large $\Gamma$ to include horizontally stacked rolls and the corresponding transitions between them. The most remarkable observation is that the transitions between the horizontal roll states have a different shape than the transitions between vertical roll states, see figure \ref{figure4}a and Ref. \cite{poe11}. The amplitudes of the $Nu$ variation over the transitions are surprisingly large, especially the transition between the SRS and the horizontal DRS. What cannot be extracted from this figure is that the lifetime of the transient states increases close to a jump. 

Figure \ref{figure4}b reveals that the ratio $f(\Gamma) \equiv n / \Gamma$ approaches a constant value of approximately $0.8$ at large $\Gamma$, where $n$ is the number of large scale rolls in the system. Here we define that for vertically stacked rolls $n \le 1$ and for horizontally stacked rolls $n \ge 1$. Comparing with $Nu(\Gamma)$ in figure \ref{figure4}a, it can be seen that there is a strong correlation between $Nu$ and $n / \Gamma$. This indicates that the asymmetric deformation of the rolls is connected to $Nu$, which can be demonstrated using the concept of plumes driving the bulk flow and transporting the heat. When a roll is horizontally elongated, the ratio between the number of rolls $n$ and $\Gamma$ decreases, this results in a lower $Nu$ as the number of thermal hotspots per horizontal plate length decreases as $1/\Gamma$ for fixed $n$. This explains the similarity of the $Nu(\Gamma)$ profiles for fixed $n$ between the $Nu(\Gamma)$ plot in figure \ref{figure4}a and the $f(\Gamma)$ plot in \ref{figure4}b. In addition, it appears that the amplitudes of the $Nu$ jumps between these plots are similar. The amplitudes decreases for each consecutive transition from $n$ to $n + 1$. This can be explained by using identical reasoning for the profile for fixed $n$, applying to a transition. We then get that the jump amplitude $\Delta Nu_n^{n+1}$ between a $n$ and $n+1$ roll state scales as

\begin{equation}
\label{deltanu}
\Delta Nu_n^{n+1} \thicksim \frac{n+1}{\Gamma} - \frac{n}{\Gamma} = \frac{1}{\Gamma}.
\end{equation}

In figure \ref{figure5}a $\Delta Nu_n^{n+1}$, extracted from $Nu(\Gamma)$ in figure \ref{figure4}a, is plotted together with a scaled function $C/(\Gamma+D)$. The offset in the denominator stems from the fact that the reasoning behind the derivation of the scaling is only valid for $n \ge 1$. The agreement between these plots is very good and therefore the fit can be used to extrapolate $\Delta Nu_n^{n+1}$ to higher $\Gamma$. Using the asymptotic values for $Nu$ and $n/\Gamma$, which are $24.92\pm0.06$ and $0.77\pm0.01$ respectively, the $\Gamma_c$ where the jump magnitude is less than 1\% of $Nu$ is found to be $\Gamma_c=26 \pm 1$, which corresponds to $n\approx20$.

Not only $Nu(\Gamma)$, but also $Re_{a}(\Gamma)$ can be used for extrapolating. Figure \ref{figure5}b shows that $Re_{a}$ increases with $\Gamma$ and approaches an asymptotic $Re_{a,\infty}$. Similar to $Nu(\Gamma)$ and for $n \ge 1$, the trend is well approximated by

\begin{equation}
\label{refit}
Re^{fit}_{a}(\Gamma) = Re_{a,\infty}(1-0.01\frac{\Gamma_c}{\Gamma}),
\end{equation}

where $\Gamma_c$ is the aspect-ratio where the function $Re^{fit}_{12}(\Gamma)$ is 99\% of $Re_{a,\infty}$. Using the $\Gamma$'s just after a jump results in $\Gamma_c = 22 \pm 1$, which is close to the prediction based on $Nu(\Gamma)$, considering the quantity used for extrapolation is very different. 

The loss of $\Gamma$ dependence is accompanied by an increase in multi-stability of the roll states. For large $\Gamma$ the transients for $\Gamma$'s near a jump in $n$ become very long-lived. As an example, for $\Gamma = 13.6$ both the $n=10$ and $n=11$ roll states survive for longer than $3000~t_E$. For comparison, a typical transient state for this $Ra = 10^8$ and $Pr = 4.3$ is less than $100~t_E$. The corresponding $Nu$ differs substantially between these roll states at $Nu = 24.68$ and $Nu = 25.11$ for $n = 10$ and $n = 11$, respectively. Thus, even though the $Nu$ dependence on $\Gamma$ decreases monotonically for larger $\Gamma$ per roll state, which can be seen in the $Nu(\Gamma)$ plot shown in figure \ref{figure4}, at large $\Gamma$ different roll states have overlapping stable regions in $\Gamma$, meaning that hysteresis effects become important. Thus, in spite of the proper extrapolations, multi-stability might prevent a solid indication for a $\Gamma_c$, at which the $\Gamma$ dependence of global properties is lost.

The effect of horizontal roll asymmetry is further highlighted by the plots in figure \ref{figure6}. The deformation indirectly results in decreasing $Nu$ by decreasing $Re_{2}$. Figure \ref{figure6}b shows that $Nu$ and $Re_2$ are positively correlated. For $\Gamma \ge 1.1$, $Re_2$ decreases monotonically with increasing $\Gamma$ due to roll asymmetry up to the $\Gamma$ where $n$ jumps to $n + 1$. In contrast, $Nu$ is negatively correlated with $Re_{1}$, see figure \ref{figure6}a, which can be explained by a flux conservation argument similar to the one presented by Grossmann and Lohse \cite{gro03}. Considering the LSC to be of simple elliptical shape, the horizontal flux has to be equal to the vertical flux. Normalizing with the height of the system, the flux balance is

\begin{equation}
g(\Gamma)\equiv \frac{Re_2}{Re_1n/\Gamma}=1.
\end{equation}

Indeed, figure \ref{figure4}c reveals that this function is in the order of $1$ over the studied $\Gamma$ range, which explains the negative correlation between $Nu$ and $Re_1$ by relating $Re_1$ to $Re_2$. 

\section{Long-lived transients} \label{Section_ICdependence}

This chapter differs from the previous two in the sense that it does not vary a single control parameter to an large extent. Instead, a phenomenon is presented in the form of very long-lived transients. Van der Poel {\it et al.} \cite{poe11} showed that for $Ra=10^7$ and $Pr=0.7$ the final state and thus $Nu$ can depend on the initial conditions of the flow field. Whether this feature of moderate Ra number RB convection can be found for high Ra number flows is an important and open question. Multiple states have been reported for high $Ra$ 3D experiments by Roche {\it et al.} \cite{roc04} and Weiss \& Ahlers \cite{wei10b}, where the system dynamically switches between two meta-stable states, each with a different $Nu$. In addition, Ahlers {\it et al.} \cite{ahl11b} showed multiple $Nu(Ra)$ scalings, dependent on minor external temperature variations. However, many details of multiple states in RB flow and their effect on $Nu$ are still unknown.

We now analyze multi-stable states at $Ra=10^9$ and $Pr=0.7$ in 2D, where the largest roll in the flow is in general significantly smaller than the limiting dimension of the enclosing boundaries. This qualification implies that simple roll states are broken down. These turbulent states with broken-down rolls are named uncondensed, where the notion of condensation is followed from \cite{xia09}. Here, a state is named condensed when the largest roll in the system is of system-size, i.e. is spectrally condensed with the energy piled-up in a scale close to system-size. In RB flow the dynamics of the coherent structures differ substantially between condensed and uncondensed states. In the uncondensed case the largest rolls are no longer stationary on a time scale shorter than a turnover time. Instead, they are mobile and cause thermal plumes to move through high-shear central regions, which is very different from the condensed states where plumes crawl along the sidewall. Whether the uncondensation is an effect of the inverse energy cascade or a result of a growing buoyancy forces compared to inertial forces or both, is not known and beyond the scope of this paper. Nevertheless, the condensation terminology is used in this chapter to highlight the similarity between condensation in two-dimensional RB flow and classical 2D \cite{kra67b} and thin layer \cite{xia09} turbulence. 

The uncondensed roll states for high $Ra$ number flows are composed of small and mobile rolls. Therefore an advanced identification algorithm is required to obtain statistics on the roll state for these cases. In the remainder of this section we first discuss the vortex detection algorithm of van der Poel {\it et al.} \cite{poe11} in detail before we apply it to a set of simulations at $Ra = 10^9$ and $Pr=0.7$, where very long-lived transients are found.

\subsection{The vortex detection algorithm}

\begin{figure*}
	\centering
\subfigure[]{\includegraphics[width=0.32\textwidth]{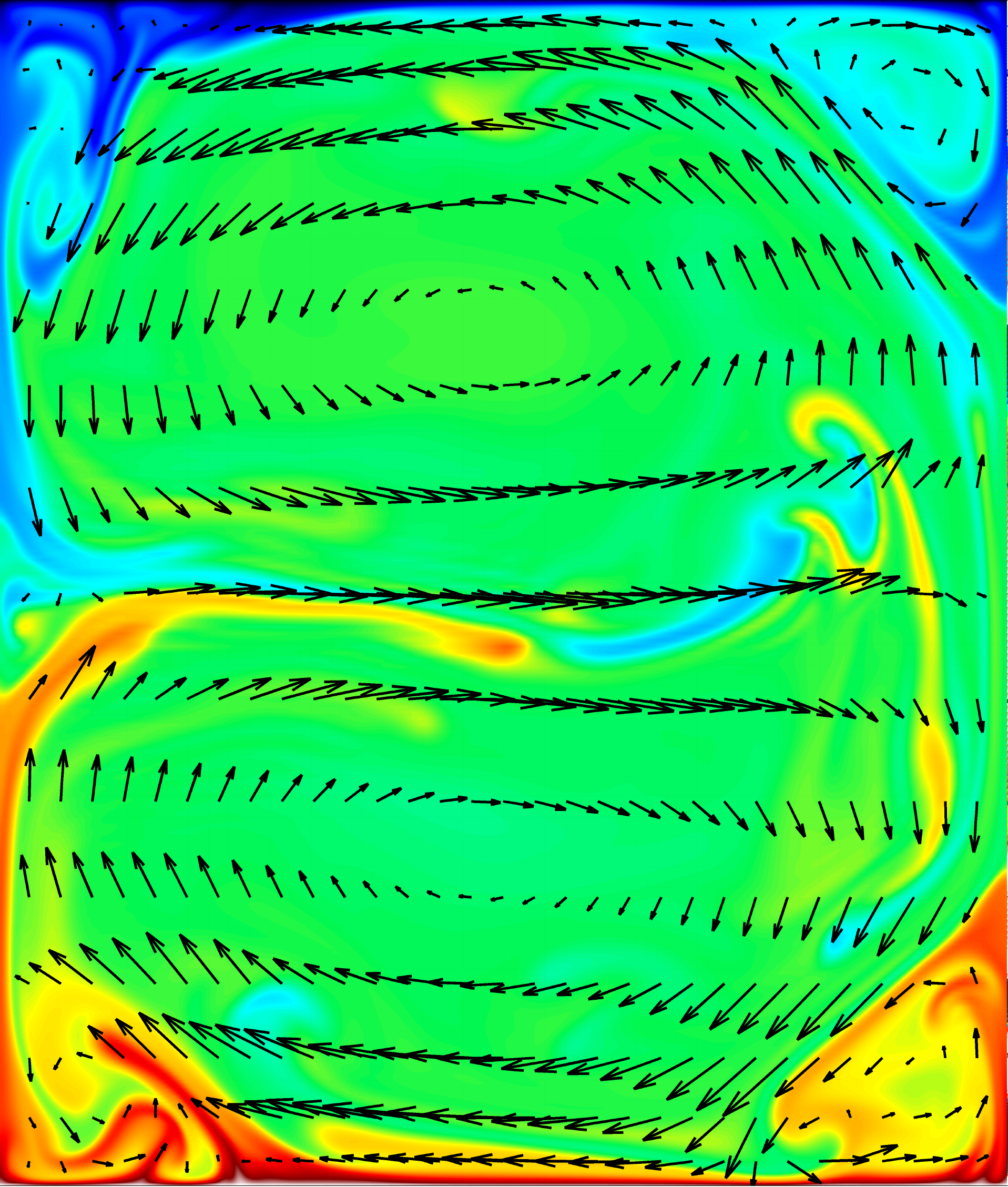}}
\subfigure[]{\includegraphics[width=0.32\textwidth]{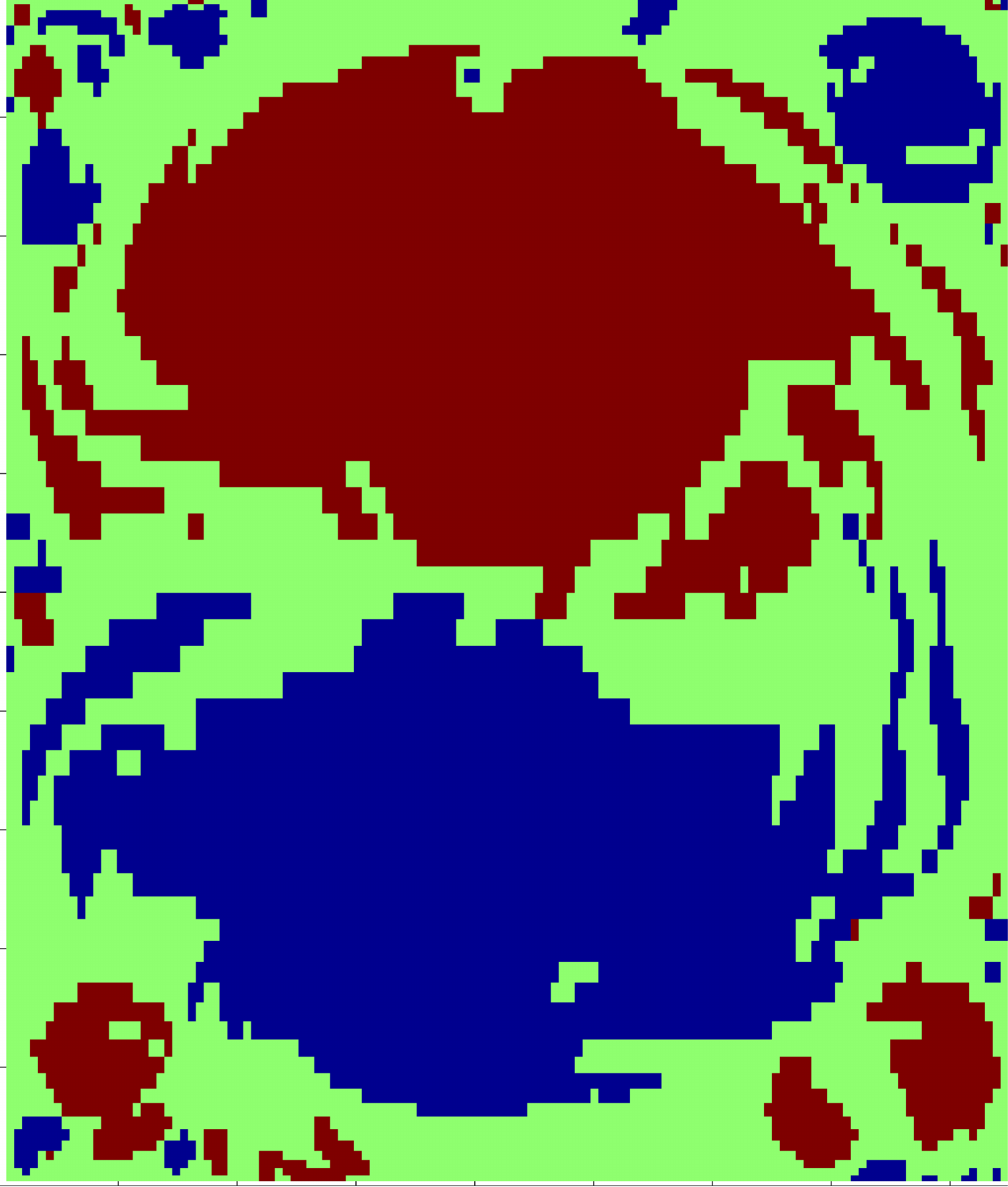}}
\subfigure[]{\includegraphics[width=0.32\textwidth]{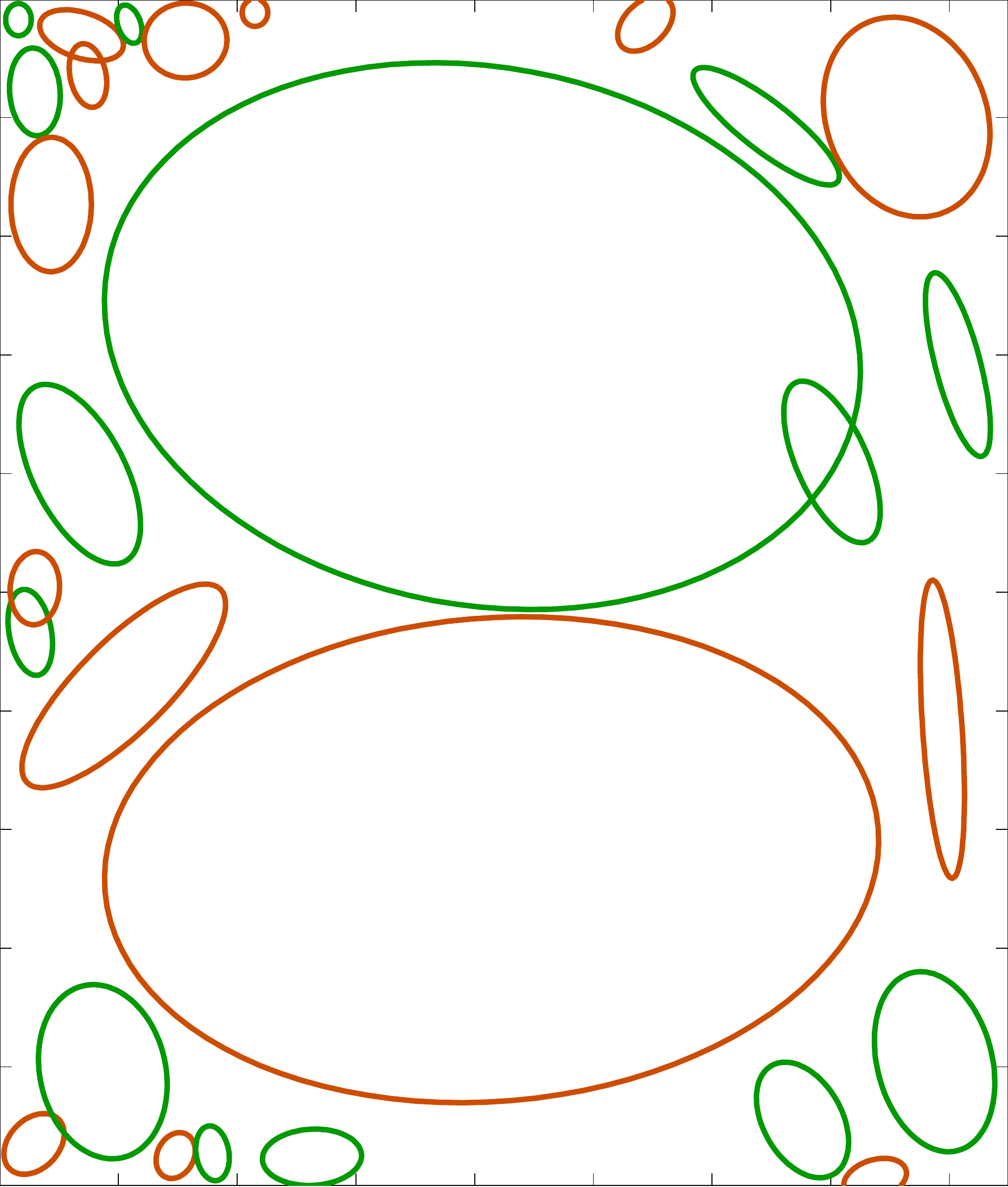}}
    \caption{Example vortex detection of a flow field snapshot for $Ra=10^9$, $Pr = 4.3$ and $\Gamma=0.85$. (a) Flow visualization with the temperature and velocity indicated by color and vectors respectively. (b) Mapped by vortex criterium and sign of vorticity. Blue and red indicate clockwise and counter clockwise rotation, respectively. Green indicates hyperbolic  field. (c) End results of the algorithm where the ellipses are colored by direction of rotation. The DRS exhibited by the flow field is captured accurately by the vortex detection algorithm.}
\label{figure7}
\end{figure*}

In order to analyze the large scale flow structure we look at the second invariant of the velocity gradient tensor $A_{ij} \equiv \partial_j u_i$. 
The criterion of a positive second invariant \cite{hun88,jeo05}, formally known as the Q-criterion, is used to identify vortex regions. For incompressible two-dimensional flow the criterion results in
\begin{equation}
A^{2}_{11} + A_{12}A_{21} < 0.
\end{equation}
A segmentation algorithm is employed to connect the grid points encompassed by the vortex regions after the grid resolution is reduced. The purpose of the grid reduction is to decrease the computational cost of the segmentation. Although some information is lost in the process, it is only detrimental for small structures, has little effect on the LSC structure and the reduction value can be adjusted. The edge points are used in a non-iterative ellipse-fit to model the vortices in the flow as ellipses. In general, this shape is more fitting for larger structures and as detecting these is the main purpose of this algorithm, it seems justified to use ellipses. The direct output consists of the six coefficients of the quadratic conic section. These coefficients are used to calculate the center coordinates, the tilt angle and the radii of the ellipse. It requires five floating points and an additional boolean for the direction of rotation, to store the information of a single vortex. The operation of the algorithm is visualized in figure \ref{figure7}. Using this method we can now store all information about the main flow structures for long simulations at $Ra=10^8$ in only $0.15$ GB, whereas storing the complete flow field over the same time-interval would require $1000$ GB. This opens the possibility to study rare events of which the information is usually lost.

\subsection{Different states in the turbulent regime}

\begin{figure*}
	\centering

\subfigure[]{\includegraphics[height=0.42\textwidth]{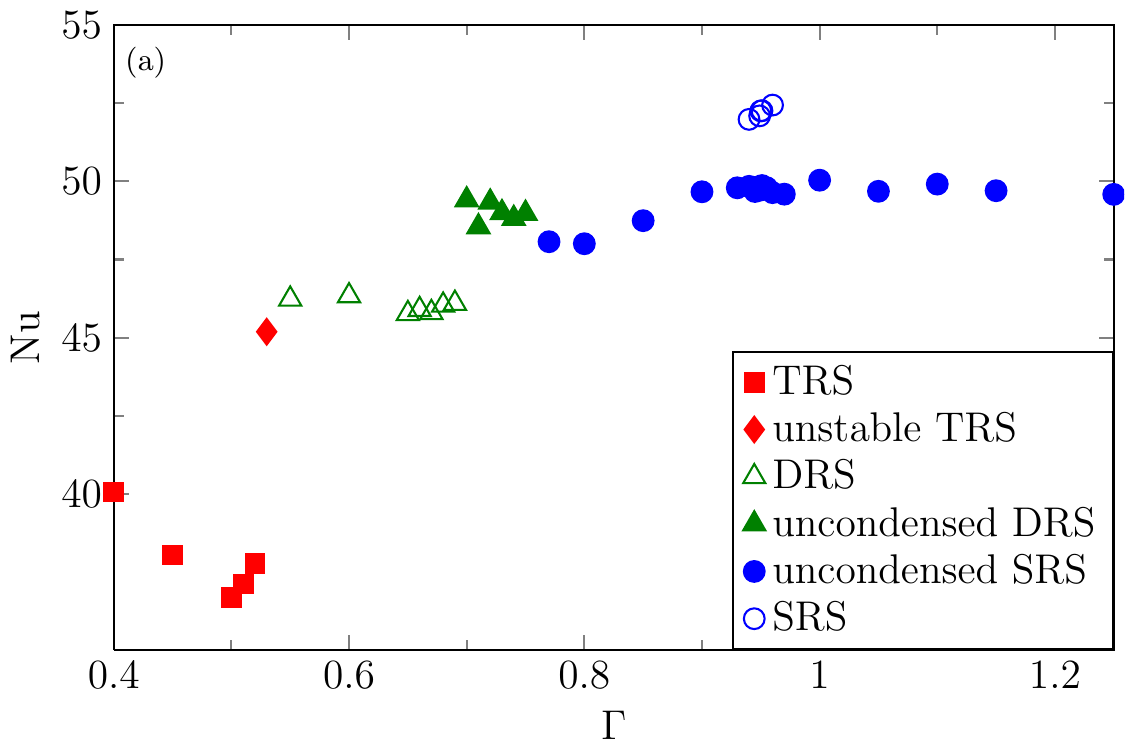}}
\subfigure[]{\includegraphics[height=0.42\textwidth]{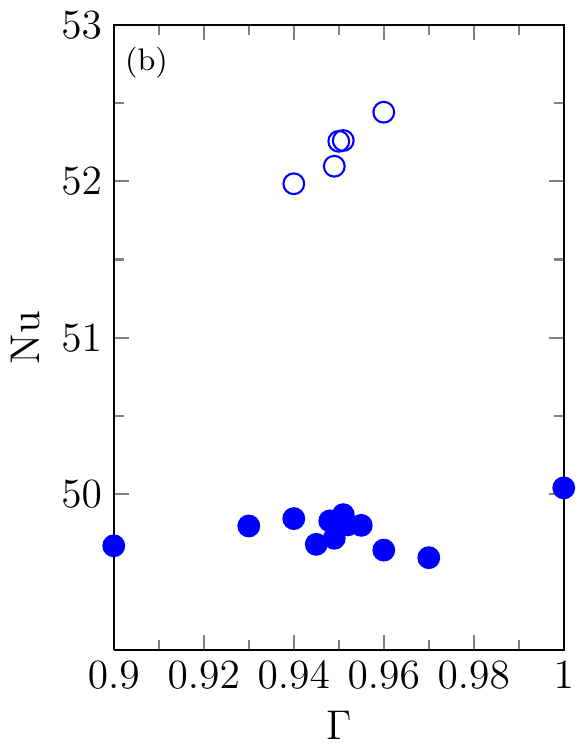}}
\subfigure[]{\includegraphics[height=0.28\textwidth]{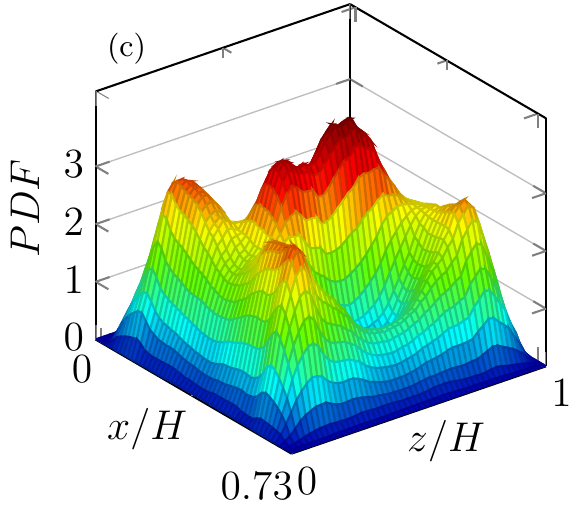}}
\subfigure[]{\includegraphics[height=0.28\textwidth]{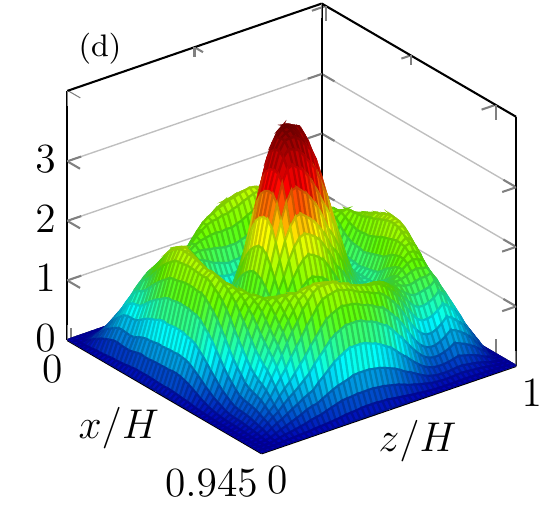}}
\subfigure[]{\includegraphics[height=0.28\textwidth]{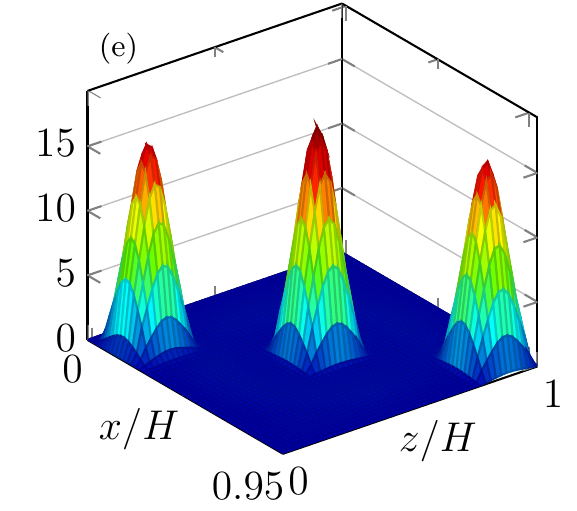}}

    \caption{(a) Nu($\Gamma$) for different states for $Pr=0.7$ and $Ra=10^9$. The error on the time-convergence is smaller than the dot size as the averaging time is more than $1000~t_E$. (b) Zoom at multi-stability region $0.90 \leq \Gamma \leq 1.00$.  2D PDF plots of the $x$ and $z$ coordinates of the vortex centroids for (c) $\Gamma=0.73$ (uncondensed DRS), (d) $\Gamma = 0.945$ (uncondensed SRS) and (e) $\Gamma = 0.95$ (SRS). In (e) the original algorithm output is convoluted with a gaussian for illustrative purpose. This results in the peaks being displayed wider than they are.}
\label{figure8}
\end{figure*}

Figure \ref{figure8} shows $Nu$ as a function of $\Gamma$. In this figure the colored symbols indicate the observed roll state using the vortex detection algorithm described above. The uncondensed SRS and uncondensed DRS states are distinct from each other by their $Nu$ and the number of equal sized largest vortices. A correlation between the roll state and $Nu$ can be distinguished as the points are clustered by Nu and the roll state. The simulation we ran for $\Gamma=0.950$ snapped into a SRS state after $750~t_E$, unlike neighboring points, which remained in the uncondensed SRS state for longer than $2200~t_E$. Forcing the system in a condensed SRS by using the steady roll state solution of $\Gamma = 0.950$ reveals that there is a possibility of multi-stability around this point as this state remains stable for longer than $1800~t_E$, see figure \ref{figure8}b. However, for $\Gamma$'s further away from $\Gamma = 0.950$, the SRS does not survive and the system immediately becomes uncondensed. This happens for e.g. $\Gamma = 1.100$, where this break down occurs within one turnover time $t_E$. Thus there appears to be hysteretic behavior not only over the jumps between the different roll states \cite{poe11}, but also over the transition to condensed states.

The SRS has a higher $Nu$ than the uncondensed SRS. This is unexpected, as flow field movies seem to reveal enhanced mixing and heat transport due to the presence of the secondary roll in the uncondensed SRS. However, a closer investigation suggests that the thermal plumes are not directly transported to the opposite boundary layer in the uncondensed SRS, which is the case for the SRS. For the TRS (triple roll state) and DRS, this effect is opposite. Here, the break up of the simple roll states increases $Nu$, similar to the effect seen in chapter III.

In contrast, the $Nu(\Gamma)$ dependence within the uncondensed SRS, uncondensed DRS and DRS regions appears to be absent. However, large $Nu$ variations can be seen at the TRS region, where due to an increased restriction of the flow, the largest scale of the flow approaches the length of the limiting dimension of the system and thereby a return of strong aspect-ratio dependence. A collection of PDFs of the $x$ and $z$ coordinates of the vortex centroids for specific and relevant roll states can be found in figure \ref{figure8}. These plots are based on the output of the vortex detection algorithm. In figure \ref{figure8}d a clear fingerprint of the 'orbiting rolls' uncondensed SRS can be observed as a large central peak with a concentric ring. The latter represents the smaller roll(s) rotating around the large central roll. Distinct from this figure is the uncondensed DRS PDF in figure \ref{figure8}c, which reveals that even different uncondensed states can be distinguished. In contrast, the SRS PDF has three sharp peaks, which are shown in figure \ref{figure8}e. The PDFs reveal that the mobility of the rolls increases for high $Ra$ uncondensed roll states, as the peaks are more smeared out for these cases. 

\section{Discussion and Conclusion}

To summarize, here we considered three cases of two-dimensional Rayleigh-B\'enard convection in direct numerical simulations. First, we showed that in small aspect ratio 2D RB convection there are two branches for the heat transport as function of $Ra$. A lower branch for relatively low $Ra$ and a higher branch for high $Ra$. By necessity, there is a strong increase of the heat transport in the transition region. This transition takes place for higher $Ra$ and is stronger when the aspect ratio is smaller. We show that the heat transport enhancement coincides with the transition from a flow state with vertically stacked stable rolls to a regime in which the thermal driving becomes strong enough to (partially) destroy the rolls. We discuss the similarities and difference with respect to the experiments of Roche {\it et al.} \cite{roc10} and Niemela $\&$ Sreenivasan \cite{nie10b} and although one cannot directly compare the 2D and 3D results in this case it is clear from this example that one should be careful in identifying an increase of the heat transport scaling as function of $Ra$ as the transition towards the ultimate regime.

Second, we showed that the influence of the aspect ratio on the heat transport decreases with increasing aspect ratio, although even at a very large $\Gamma=10$ variations up to $2.5\%$ are observed due to changes in the flow structure. Extrapolating $Re_{a}(\Gamma)$ towards high $\Gamma$ gives $\Gamma_c = 22 \pm 1$, where aspect-ratio dependence of global properties is less than 1\% of their asymptotic values. Using the jumps in $Nu(\Gamma)$, a value of $\Gamma_c=26 \pm 1$ is found. 

Finally, we observe very long-lived transients / different turbulent states at $Ra=10^8$ and $Pr=4.3$ for certain very large aspect ratios and at $Ra=10^9$ and $Pr=0.7$. In the latter regime rare random events cause the system to switch between states. The probability of these events occurring, the stability of the resulting state and the effect on $Nu$ appear to be very sensitive to the control parameters. In particular, the aspect-ratio $\Gamma$ has a large influence on these phenomena; varying $\Gamma$ reveals a rich phase space for a given $Ra$ and $Pr$. For these parameters the flow is so turbulent that the different states with their rare events can only be identified with the use of an advanced vortex detection algorithm, which has revealed the possibility of hysteretic behavior over the transition to condensed states. 

\vspace{0.5cm}
\noindent \textbf{Acknowledgements:} The work was supported by the Foundation for Fundamental Research on Matter (FOM) and the National Computing Facilities (NCF), both sponsored by NWO. The computations have been performed on the LISA cluster of SARA.

\end{document}